\documentclass[final,5p,times]{elsarticle}

\usepackage{graphicx}
\graphicspath{{figures/}}
\usepackage{color}
\usepackage[nodots]{numcompress}
\usepackage{microtype}
\usepackage{fixltx2e}
\usepackage{mathtools}
\usepackage{flushend}
\usepackage[pdftitle={New Transforms for JPEG Format},pdfauthor={Stanislav Svoboda, David Barina},pdfsubject={SCCG 2017}]{hyperref}

\makeatletter
\providecommand{\doi}[1]{%
  \begingroup
    \let\bibinfo\@secondoftwo
    \urlstyle{rm}%
    \href{http://dx.doi.org/#1}{%
      doi:\discretionary{}{}{}%
      \nolinkurl{#1}%
    }%
  \endgroup
}
\makeatother

\journal{SCCG 2017}

\begin{document}

\begin{frontmatter}

\title{New Transforms for JPEG Format}
\author{Stanislav Svoboda}
\ead{xsvobo0b@stud.fit.vutbr.cz}
\author{David Barina\corref{cor1}}
\ead{ibarina@fit.vutbr.cz}
\address{Faculty of Information Technology\\
Brno University of Technology \\
Bozetechova 1/2 \\
Brno, Czech Republic}
\cortext[cor1]{Corresponding author}

\begin{abstract}
The two-dimensional discrete cosine transform (DCT) can be found in the heart of many image compression algorithms.
Specifically, the JPEG format uses a lossy form of compression based on that transform.
Since the standardization of the JPEG, many other transforms become practical in lossy data compression.
This article aims to analyze the use of these transforms as the DCT replacement in the JPEG compression chain.
Each transform is examined for different image datasets and subsequently compared to other transforms using the peak signal-to-noise ratio (PSNR).
Our experiments show that an overlapping variation of the DCT, the local cosine transform (LCT), overcame the original block-wise transform at low bitrates.
At high bitrates, the discrete wavelet transform employing the Cohen--Daubechies--Feauveau 9/7 wavelet offers about the same compression performance as the DCT.
\end{abstract}


\end{frontmatter}

\section{Introduction}
\label{sec:Introduction}

\begin{figure*}
	\def\svgwidth{\linewidth}
	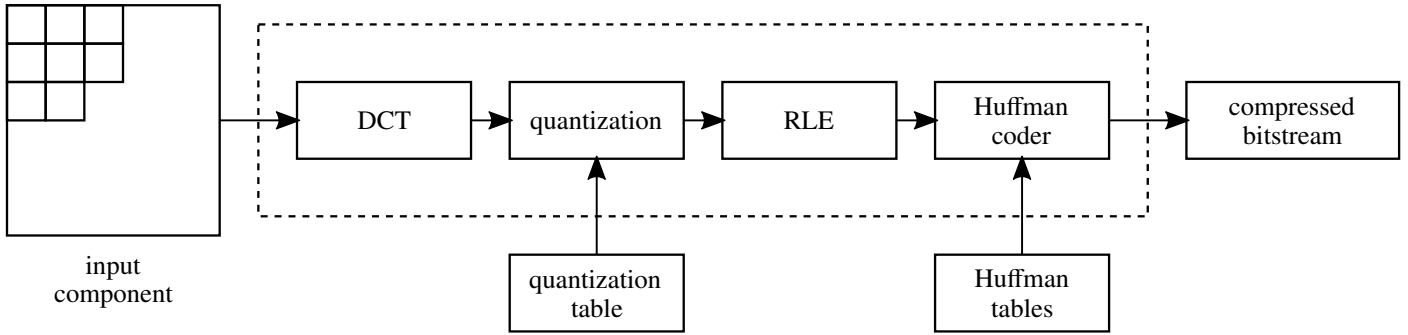
	\caption{
		JPEG overview.
		The dashed line indicates the compression chain.
	}
	\label{fig:JPEG-overview}
\end{figure*}

In last decades, needs for high-quality photography are growing, and so demands for efficient data storage are also growing.
Therefore, it is important to compress the data as much as possible while preserving the quality of the image.
For example, transferring a large number of images with high resolution across the Internet without a certain level of compression would be very time-consuming.
Regarding the photography, the problem can be addressed by lossy image compression.
Nowadays, the JPEG standard \cite{std:JPEG}, dating back to 1991, is still the most widely used format for the lossy compression.
Figure~\ref{fig:JPEG-overview} shows the underlying compression chain.
Each color component is transformed by blocks $8\times8$ using the \mbox{2-D} DCT.
The DCT has the property that, for a typical image, most of the visually significant information about the image in the $8\times8$ block is concentrated in just a few coefficients.
This allows for better image compression.
The transform coefficients are further quantized and fed into an entropy coder.

Since then, several other lossy image compression standards have been standardized.
However, none of them became more popular with the public than the original JPEG.
Particularly, the JPEG 2000 \cite{std:JPEG2000} decomposes large image tiles using the discrete wavelet transform (DWT).
The advantage of wavelets is that wavelets are located on the small area in the image domain.
Another interesting standard is JPEG XR \cite{std:JPEGXR}, which is based on an overlapping hierarchical transform, so-called lapped biorthogonal transform (LBT).
The last of the standards to be mentioned is WebP \cite{std:WebP}, based on the DCT complemented by Walsh--Hadamard transform (WHT).

Figure~\ref{fig:JPEG-overview} shows that the JPEG block-wise scheme is very general.
This opens the way to incorporate some other suitable transforms into the same compression chain.
This is the motivation behind our research.

The rest of the paper is organized as follows.
Section~\ref{sec:JPEG Format} presents the JPEG chain in the necessary level of detail.
Subsequent Section~\ref{sec:New Transforms for JPEG Format} deals with the transforms suitable for involvement in this chain and examines their compression capabilities.
Eventually, Section~\ref{sec:Conclusions} summarizes and closes the paper.

\section{JPEG Format}
\label{sec:JPEG Format}

Part 1 of JPEG standard \cite{std:JPEG} specifies the method of lossy compression for digital images, based on the discrete cosine transform (DCT).
This section describes a simplified description of JPEG image compression.

The color model to be used is YC\textsubscript{B}C\textsubscript{R}.
Therefore, the representation of the colors in the image is first converted from RGB to YC\textsubscript{B}C\textsubscript{R}.
The transformation into the YC\textsubscript{B}C\textsubscript{R} model enables the next usual step, which is to reduce the spatial resolution of the C\textsubscript{B} and C\textsubscript{R} components.
For the rest of the compression process, Y, C\textsubscript{B}, and C\textsubscript{R} components are processed separately.

\begin{figure}[h]
	\centering
	\includegraphics[width=.49\linewidth]{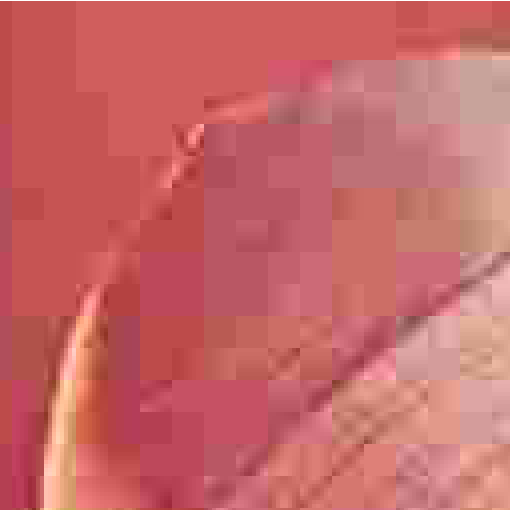}
	\caption{Blocking artifacts caused by the JPEG compression.}
	\label{fig:blocking-artifacts}
\end{figure}

As a next step, each component is split into blocks of $8\times8$ samples, $x_{k,l}$ for $(0,0) \le (k,l) < (8,8)$.
The samples are then shifted down by 128, assuming an \mbox{8-bit} depth.
Subsequently, each block undergoes the two-dimensional discrete cosine transform
\begin{align}
	X_{m,n} = \lambda_{m,n} \sum_{k,l} \cos\left(\frac{\pi{}(k+1/2)m}{N}\right) \cos\left(\frac{\pi{}(l+1/2)n}{N}\right) \, x_{k,l} \text{,}
\end{align}
where $(0,0) \le (m,n) < (8,8)$, $\lambda_{m,n}$ is a scaling factor, and $N=8$.
Now, the amplitudes of the coefficients are quantized.
When performing a block-based transform and quantization, several types of artifacts can appear, especially blocking artifacts.
The blocking artifacts are shown in Figure~\ref{fig:blocking-artifacts}.
The artifacts can be reduced by choosing a finer quantization, which corresponds to a lower level of compression.

The DCT itself is a lossless process since the original input can be exactly reconstructed by applying an inverse transform to the coefficients $X_{m,n}$ directly.
In order to achieve substantial compression ratio, quantization is applied to reduce the levels of the coefficients.
The uniform quantization procedure is used to quantize the coefficients.
One of up to four quantization tables $Q_{m,n}$ may be used in the quantization procedure.
No default quantization table is specified in the standard.
The quantization is formulated as
\begin{align}
	\hat{X}_{m,n} = \operatorname{round} \left( \frac{X_{m,n}}{Q_{m,n}} \right) \text{,}
\end{align}
where the $\operatorname{round}(a)$ operator rounds value $a$ to the nearest integer.
Human visual system is more immune to a loss of higher spatial frequency components than a loss of lower frequency components.
This allows quantization to greatly reduce the amount of information in the high-frequency components.

After quantization, the $\hat{X}_{m,n}$ coefficients are fed into an entropy coder.
Entropy coding employed in the JPEG is a special form of lossless compression.
The $\hat{X}_{0,0}$ coefficient (DC coefficient) is treated differently than other coefficients (AC coefficients).
The latter ones are converted into a one-dimensional "zig-zag" sequence.
The rest of the process involves run-length encoding (RLE) of zeros and then using Huffman coding (arithmetic coding is possible, however rarely used).

From the above, it is clear that the scheme is almost independent of the transform used.
Consequently, it would seem logical to substitute the DCT with some other similar transform.
Several other papers on this topic have already been published.
Some of them are briefly reviewed below.
The authors of \cite{Mukundan2006} examined the possibility of using the discrete Chebyshev transform (DChT) in JPEG.
As reported in their paper, the DChT overcomes DCT on images with sharp edges and high predictability.
In \cite{Malvar1997}, the author compared the compression performance of the block-wise DCT against several lapped transforms.
He concluded that lapped transforms have less blocking than the DCT and show some PSNR improvement over the DCT.

Considering the existing papers, we see that a wider comparison of the transforms in the JPEG compression chain is missing.
The next section investigates the performance of some promising transforms in conjunction with the JPEG compression.

\section{New Transforms for JPEG Format}
\label{sec:New Transforms for JPEG Format}

This section interleaves a description of the transforms and their evaluation.
The evaluation was performed on two datasets \cite{Olmos2004,Franzen}.
At the beginning, trigonometric transforms are investigated.
Subsequently, separable and non-separable wavelet, Chebyshev, and Walsh--Hadamard transforms are examined.

\subsection{Discrete Sine Transform}

The discrete sine transform (DST) is very similar to the DCT, except cosines are replaced with sines.
Recall that the DCT has the property that, for a typical image, most of the information is concentrated in just a few coefficients $X_{m,n}$ with the lowest $({m,n})$ indices.
However, this property is not always valid for sine transforms.
We found one variant for which the property holds.
In the literature, this variant is referred to as the \mbox{DST-VII} \cite{Puschel2003} variant.
Since most of the transforms investigated in this paper are separable, only the one-dimensional definitions are given from now on.
The DST is defined by
\begin{align}
	X_m = \lambda_m \sum_k \sin\left( \frac{\pi{}(k+1)(m+1/2)}{N+1/2} \right) \, x_k \text{,}
\end{align}
where $\lambda_{m}$ is a scaling factor.

\subsection{Discrete Hartley transform}

Like the previous transforms, the discrete Hartley transform (DHT) \cite{Bracewell1983} is also based on trigonometric functions.
In fact, its definition looks very similar to the definition of discrete Fourier transform (DFT).
Unlike the DFT, the discrete Hartley transform maps real inputs onto real outputs, with no involvement of complex numbers.
The transform is defined by
\begin{align}
	X_m = \sum_k \operatorname{cas}\left( \frac{2\pi{}km}{N} \right) \, x_k \text{,}
\end{align}
where $\operatorname{cas}(\alpha) = \cos(\alpha) + \sin(\alpha)$.

\subsection{Local Cosine Transform}

The local cosine transform (LCT) \cite{Aharoni1993} reduces and smoothes the block effects.
The local cosine transform is based on the standard block-based DCT.
However, the local cosine transform has basis functions that overlap adjacent blocks.
Prior to the DCT algorithm, a preprocessing phase in which the image is multiplied by smooth bell functions that overlap adjacent blocks is applied.
This phase is implemented by folding the overlapping parts of the bells back into the original blocks.
The standard block-based DCT algorithm then operates on the resulting blocks.

The folding operations are defined as
\begin{align}
	f_{-}(n) = \frac{b(n)f(-n)-b(-n)f( n)}{b(n)-b(-n)} \text{,} \\
	f^{+}(n) = \frac{b(n)f( n)-b(-n)f(-n)}{b(n)-b(-n)} \text{,}
\end{align}
where the $f_{-}(n)$ is $n$th coefficient to the left (top) of the current block,
the $f^{+}(n)$ is $n$th coefficient to the right (bottom),
and $b(n) = \beta( (2n+1)/N )$ is a bell function, where
\begin{align}
	\beta(n) = \begin{cases}
		0 & n < -1 \\
		\frac{1+\sin(\pi{}n/2)}{2} & \text{otherwise} \\
		1 & n > +1
	\end{cases} \text{.}
\end{align}

\begin{figure}
	\centering
	\includegraphics[width=.8\linewidth]{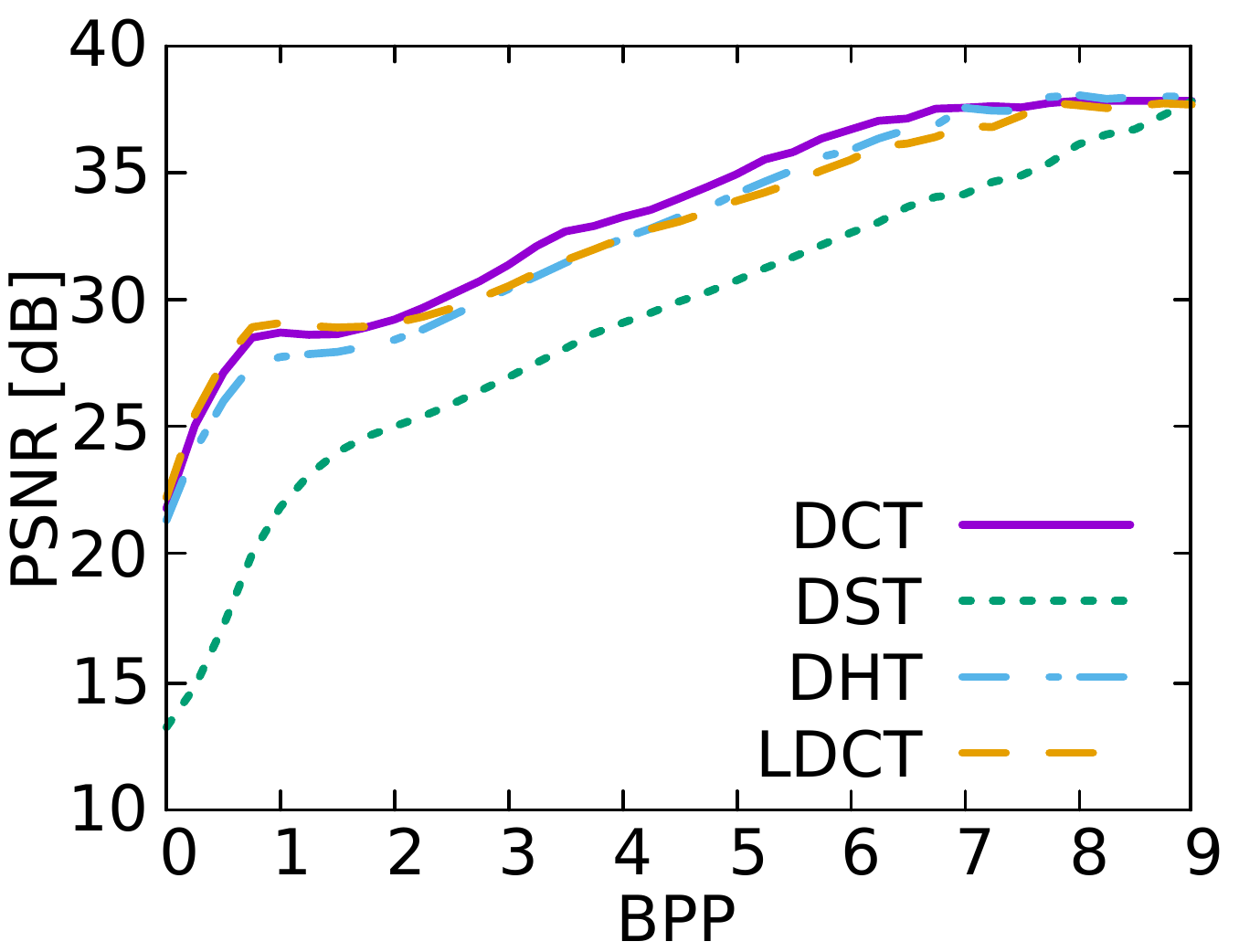}
	\caption{
		Comparison of the DCT, DST, DHT, and LCT.
		The LCT overcomes the other transforms at low bitrates.
	}
	\label{fig:psnr-trigonometric}
\end{figure}

\begin{figure}
	\centering
	\includegraphics[width=\linewidth]{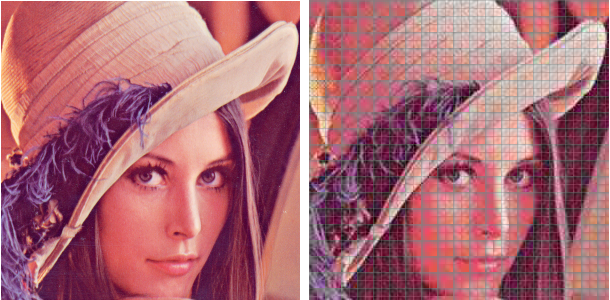}
	\caption{
		Sample image (on the left) and DST artifacts on block boundaries (on the right).
	}
	\label{fig:dst-artifacts}
\end{figure}

The comparison of all the transforms discussed above is shown in Figure~\ref{fig:psnr-trigonometric}.
The $x$-axis indicates bits per pixel (bpp).
The discrete sine transform performs significantly worse than the reference DCT.
This is caused by artifacts on block boundaries, as shown in Figure~\ref{fig:dst-artifacts}.
Also, the discrete Hartley transform performs worse than the DCT.
As we have found, this is caused by blocking artifacts at higher bitrates, where the artifacts are no longer visible with the DCT.
At lowest bitrates, the local cosine transform has a better image quality than the reference DCT.
Unfortunately, at higher bitrates, the image quality is slightly worse.
The results on lower bitrates are caused by reduced blocking artifacts.

\subsection{Discrete Wavelet Transform}

The discrete wavelet transform (DWT) became a very popular image processing tool in last decades.
For example, the JPEG 2000 standard is based on such decomposition technique.
In more detail, the DWT decomposes the image into several subbands, while employing simple basis functions, the wavelets \cite{Daubechies1992}.
The transform is usually applied on large image tiles instead of small $8\times8$ blocks.
Consequently, there are no blocking artifacts at all.
In this paper, two well-known biorthogonal wavelets are used, the Cohen--Daubechies--Feauveau (CDF) \cite{Cohen1992} 5/3 and 9/7 wavelets.
Incidentally, both of them are employed in the JPEG 2000 standard.
In order to fit into JPEG compression chain, the wavelet transforms were designed to create a regular $8\times8$ grid of coefficients.
This design corresponds to three levels of a dyadic decomposition \cite{Mallat1989}.
The coordinates of the coefficients in the $8\times8$ blocks are computed using bit-reversal operations.
In this way, the coefficients more closely correspond to the DCT coefficients.
Note that both of the transforms were implemented using a lifting scheme \cite{Daubechies1998,Sweldens1996}.
The lifting scheme can decompose the wavelet transforms into a finite sequence of simple filtering steps (lifting steps).
Usually, the first step in the pair is referred to as the predict and the second one as the update.

\subsection{Red-Black Wavelet Transform}

The red-black wavelet transform (DWT RB) \cite{Uytterhoeven1998} is computed using a \mbox{2-D} lifting scheme on a quincunx lattice \cite{Feilner2005}.
The wavelets constructed in this way are inherently non-separable.
Consequently, the red-black wavelets are less anisotropic than the classical tensor product wavelets (the classical DWT).
In other words, the classical DWT will favor horizontal and vertical features of the image.
This is not visible in the red-black wavelet transform.

\begin{figure}
	\centering
	\includegraphics[width=.8\linewidth]{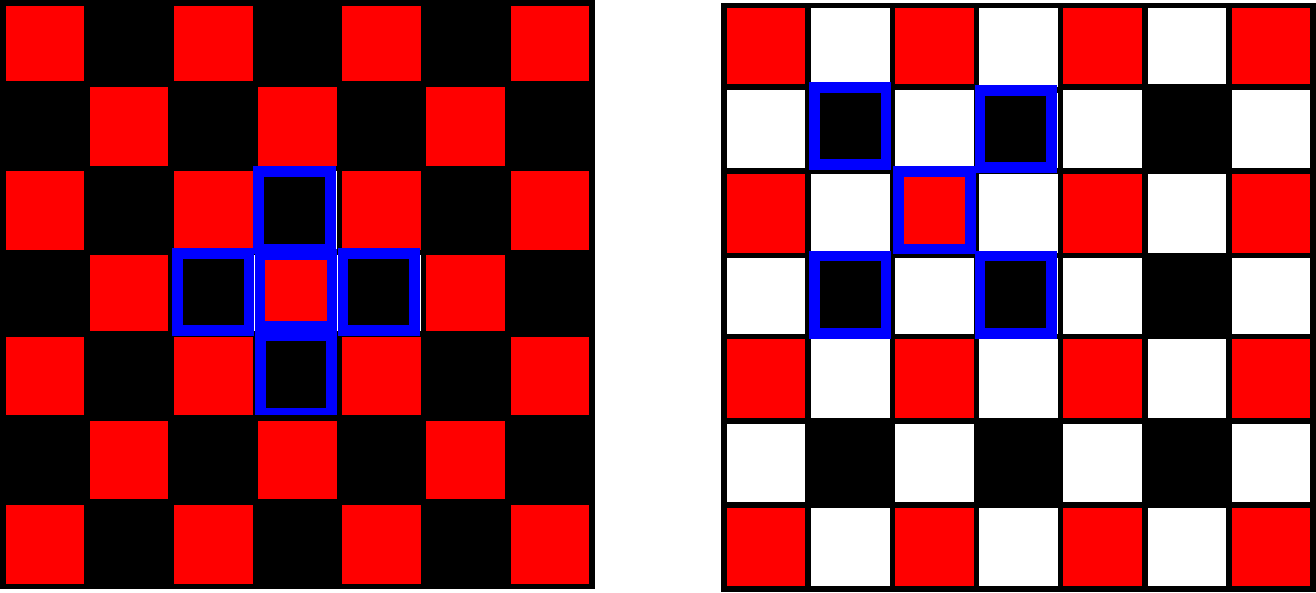}
	\caption{
		Lifting scheme on the quincunx lattice.
		Filter samples are bordered in blue.
		The first step (predict) on the left, the second (update) on the right.
	}
	\label{fig:lifting-quincunx}
\end{figure}

The construction of the red-black wavelets is based on the CDF wavelets above.
Therefore, the CDF 5/3 and CDF 9/7 wavelets are used also for this construction.
The individual steps of the lifting scheme are illustrated in Figure~\ref{fig:lifting-quincunx}.
The details can be found in \cite{Uytterhoeven1998}.

\begin{figure}
	\centering
	\includegraphics[width=.8\linewidth]{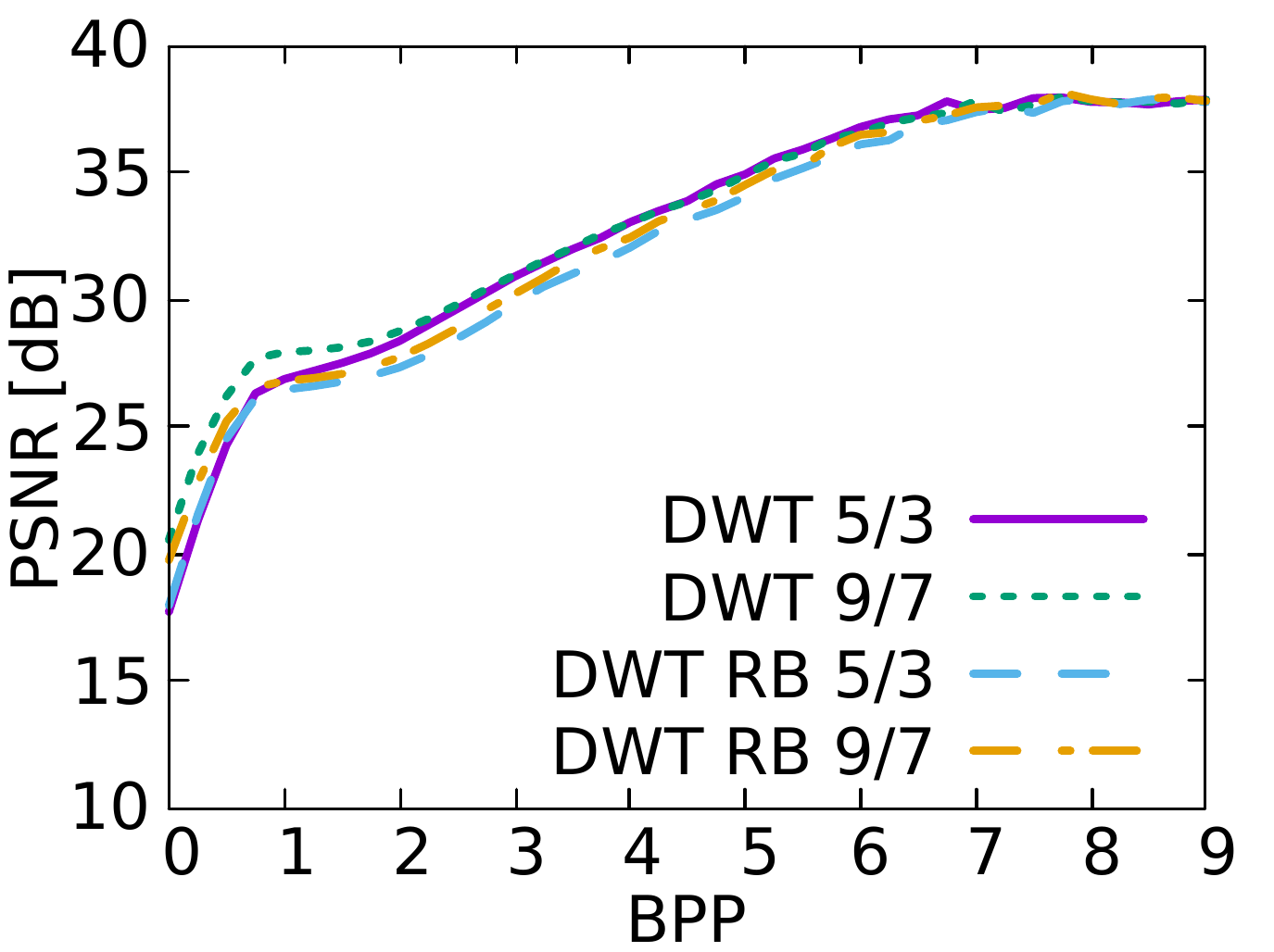}
	\caption{
		Comparison of the separable and non-separable (red-black) wavelet transforms.
		The separable CDF 9/7 transform has the best overall performance.
	}
	\label{fig:psnr-wavelets}
\end{figure}

\begin{figure}
	\centering%
	\includegraphics[width=0.49\linewidth]{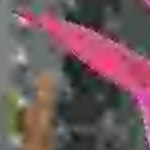}%
	\hfill%
	\includegraphics[width=0.49\linewidth]{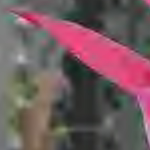}%
	\caption{
		Comparison of the non-separable CDF 9/7 (on the left) wavelet transform and separable CDF 9/7 (on the right) in artefacts at edges.
	}
	\label{fig:wavdiff}
\end{figure}

According to the results shown in Figure~\ref{fig:psnr-wavelets}, it is clear that the separable CDF transform performs better than the red-black transforms.
The worse results of the non-separable CDF 9/7 are caused by stain artifacts on the edges of objects, as illustrated in Figure~\ref{fig:wavdiff}.
The CDF 9/7 transforms perform always better than the CDF 5/3.
Therefore, the best combination seems to be separable the CDF 9/7 transform.

\subsection{Discrete Chebyshev Transform}

The discrete Chebyshev transform (DChT) \cite{Corr2000} is a polynomial-based transform, which employs Chebyshev polynomials of the first kind $T_n(x)$.
Since the DCT is closely associated with a Chebyshev Polynomial series as $\cos(n\alpha)=T_n(\cos(\alpha))$ for some $\alpha$, the discrete Chebyshev transform can be viewed as a natural modification of the DCT.
The discrete Chebyshev transform is then defined using the polynomials
\begin{align}
	t_p(x) = (A_1x+A_2)t_{p-1}(x) + A_3t_{p-1}(x) \text{,}
\end{align}
where $A_1\text{, } A_2\text{,}$ and $A_3$ are constant.
The transform is then defined as
\begin{align}
	X_m = \sum_k t_m(k) \, x_k \text{.}
\end{align}

\subsection{Walsh--Hadamard Transform}

The last of the transforms discussed in this paper is the Walsh--Hadamard transform (WHT).
The computation \cite{Shanks1969} of this transform should be very fast since only additions/subtractions are involved here.
The transform is defined as
\begin{align}
	X_m = 1/N \sum_k W(m,k) \, x_k \text{,}
\end{align}
where
\begin{align}
	W(m,k) = \begin{bmatrix}
		+1 & +1 & +1 & +1 & +1 & +1 & +1 & +1 \\
		+1 & +1 & +1 & +1 & -1 & -1 & -1 & -1 \\
		+1 & +1 & -1 & -1 & +1 & +1 & -1 & -1 \\
		+1 & +1 & -1 & -1 & -1 & -1 & +1 & +1 \\
		+1 & -1 & +1 & -1 & +1 & -1 & +1 & +1 \\
		+1 & -1 & +1 & -1 & -1 & +1 & -1 & -1 \\
		+1 & -1 & -1 & +1 & +1 & -1 & -1 & +1 \\
		+1 & -1 & -1 & +1 & -1 & +1 & +1 & -1
	\end{bmatrix} \text{.}
\end{align}

\begin{figure}[b]
	\centering
	\includegraphics[width=.8\linewidth]{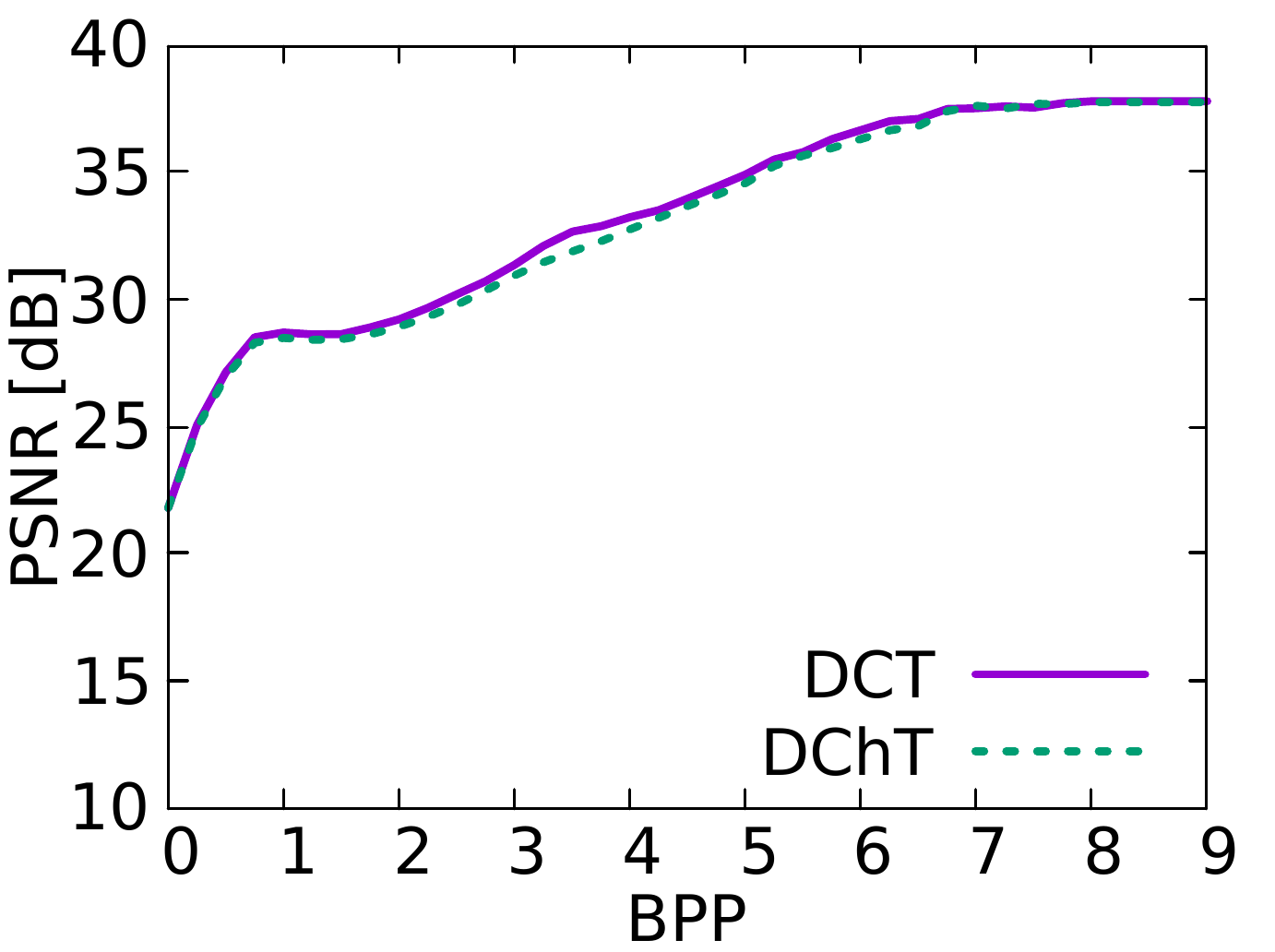}
	\caption{
		Comparison of the DCT and DChT.
		The DCT is slightly better.
	}
	\label{fig:psnr-polynomial}
\end{figure}

\begin{figure}
	\centering
	\includegraphics[width=.8\linewidth]{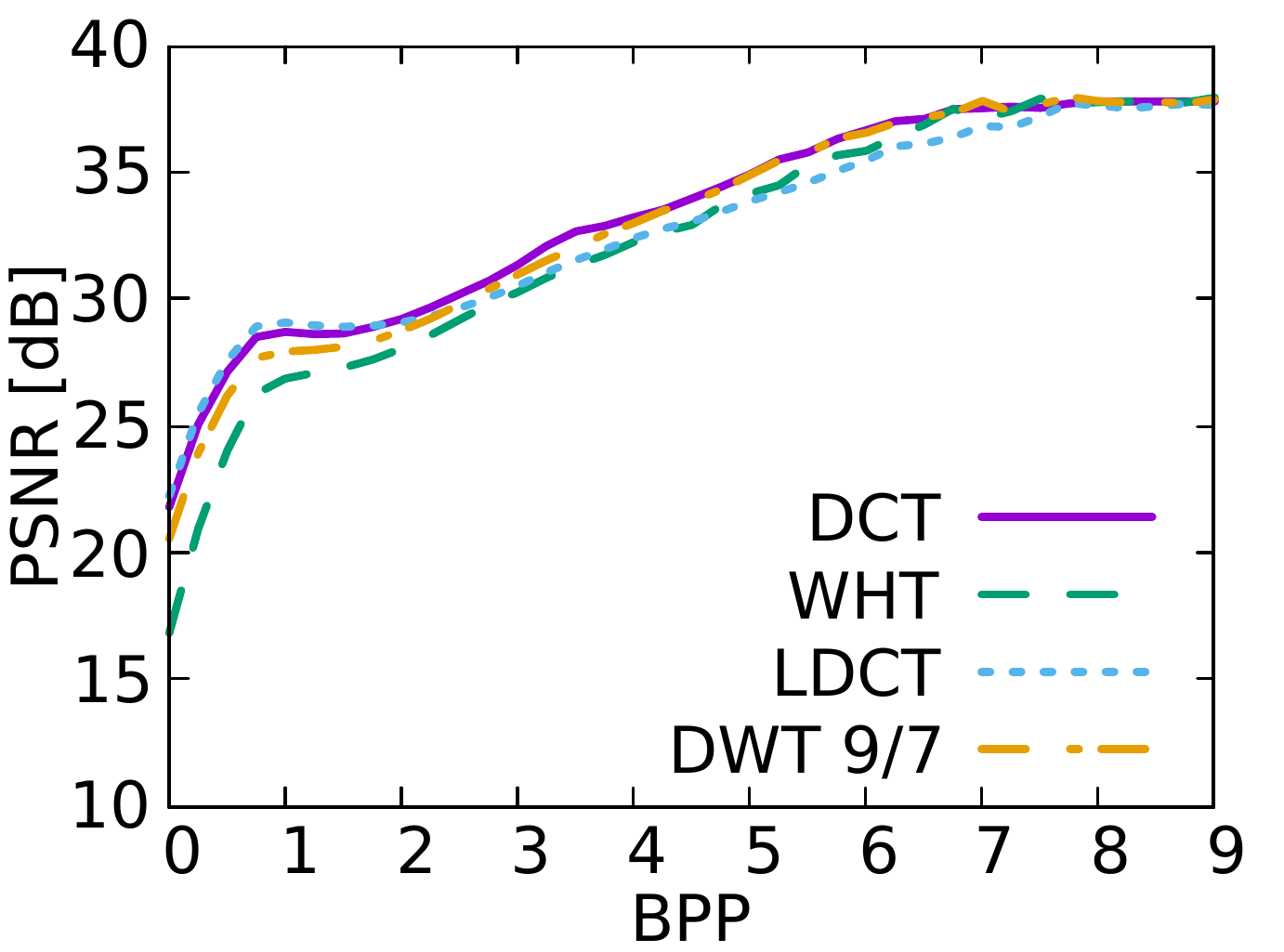}
	\caption{
		Overall comparison of selected transforms.
	}
	\label{fig:psnr-overall}
\end{figure}

\begin{figure*}
	\centering%
	\hfill%
	\includegraphics[width=.3\linewidth]{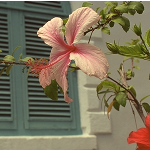}%
	\hfill%
	\includegraphics[width=.3\linewidth]{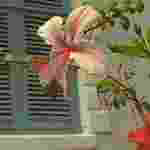}%
	\hfill%
	\includegraphics[width=.3\linewidth]{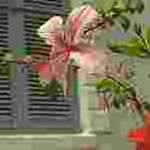}%
	\hfill\null%
	\caption{
		Visual comparison of the original image (left), DCT (middle, PSNR 23.0 dB), and LCT (right, PSNR 23.3 dB).
	}
	\label{fig:strip-dct-lct}
\end{figure*}

\begin{figure*}
	\centering
	\hfill%
	\includegraphics[width=.24\linewidth]{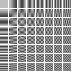}%
	\hfill%
	\includegraphics[width=.24\linewidth]{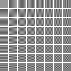}%
	\hfill%
	\includegraphics[width=.24\linewidth]{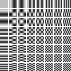}%
	\hfill%
	\includegraphics[width=.24\linewidth]{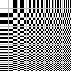}%
	\hfill\null%
	\caption{
		Basis images of selected transforms, from the left: the DCT, DChT, DHT, and WHT.
		DC coefficient is located in the top left corner.
	}
	\label{fig:bases}
\end{figure*}

\begin{figure*}
	\centering%
	\hfill%
	\includegraphics[width=.3\linewidth]{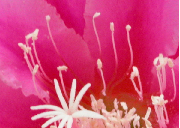}%
	\hfill%
	\includegraphics[width=.3\linewidth]{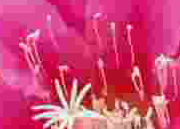}%
	\hfill%
	\includegraphics[width=.3\linewidth]{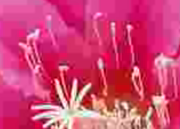}%
	\hfill\null%
	\caption{
		Blocking artifacts, from the left: the original image, DCT, and LCT.
	}
	\label{fig:strip-dct-lct-2}
\end{figure*}

Figure~\ref{fig:psnr-polynomial} shows the performance of the discrete Chebyshev transform.
It is evident that the DChT is everywhere slightly below the DCT.
Finally, Figure~\ref{fig:psnr-overall} shows an overall comparison, including the Walsh--Hadamard transform.
Also, the WHT does not overcome DCT in any part of the plot.
The only advantage of the WHT is the computation performance.
In summary, only the local cosine transform (LCT) overcame the original block-wise DCT, especially at low bitrates.
In addition, it removes blocking artifacts, as documented in Figures~\ref{fig:strip-dct-lct} and \ref{fig:strip-dct-lct-2}.
Additionally, the separable discrete wavelet transform with the CDF 9/7 wavelet offers about the same compression performance as the DCT at high bitrates.
To improve the mental image of evaluated transforms, bases of selected transforms are visually compared in Figure~\ref{fig:bases}.

\section{Conclusions}
\label{sec:Conclusions}

The JPEG image compression format uses a lossy form of compression based on the discrete cosine transform.
This paper deals with a substitution the discrete cosine transform in the JPEG compression with some other similar transform.
Several practical transforms were examined, including other trigonometric transforms, separable and non-separable wavelet transforms, a transform employing Chebyshev polynomials, and the Walsh--Hadamard transform.
These transforms were evaluated on several image datasets.

The experiments show that only the local cosine transform overcomes the original block-wise DCT at low bitrates.
Besides, it removes blocking artifacts.
At high bitrates, the CDF 9/7 discrete wavelet transform performs similarly as the DCT.

In future work, we plan to focus on other transforms that have not been covered here, including also non-linear transforms.

\subsection*{Acknowledgements}

This work has been supported by
the Ministry of Education, Youth and Sports from the National Programme of Sustainability (NPU II) project IT4Innovations excellence in science (no. LQ1602), and
the Technology Agency of the Czech Republic (TA CR) Competence Centres project V3C -- Visual Computing Competence Center (no. TE01020415).

\bigskip
\null

\section*{References}
\bibliographystyle{model3-num-names}
\bibliography{sources}

\end{document}